\documentclass[onecolumn]{emulateapj}

\usepackage{lscape}
\usepackage{amsmath}
\usepackage{xspace}
\usepackage{graphicx}

\newcommand{\sax}{{\it BeppoSAX} }
\newcommand{\asca}{{\it ASCA} }

\newcommand{\exosat}{{\it EXOSAT} }
\newcommand{\einstein}{{\it Einstein} }
\newcommand{\integral}{{\it INTEGRAL} }

\newcommand{\temna}{{\it Temna} }

\newcommand{\source}{4U~1636--53 }

\begin{document}

%% LaTeX will automatically break titles if they run longer than
%% one line. However, you may use \\ to force a line break if
%% you desire.

\title{Disc-Jet coupling in the LMXB \source from \integral\/ 
{\thanks{INTEGRAL is an ESA project with instruments 
 and science data centre funded by ESA member states 
(especially the PI countries: Denmark, France, Germany, Italy, Switzerland, Spain),
 Czech Republic and Poland, and with the participation of Russia and the USA.}} 
observation\\
}

%% Use \author, \affil, and the \and command to format
%% author and affiliation information.
%% Note that \email has replaced the old \authoremail command
%% from AASTeX v4.0. You can use \email to mark an email address
%% anywhere in the paper, not just in the front matter.
%% As in the title, use \\ to force line breaks.

\author{Mariateresa  Fiocchi, Angela Bazzano, Pietro Ubertini}
\affil{Istituto di Astrofisica Spaziale e Fisica Cosmica di Roma (INAF)\\ Via Fosso del Cavaliere 100, Roma, I-00133, Italy} 
%\author{Andrzej Zdziarski}
%\affil{Nicolaus Copernicus Astronomical Center, Bartycka 18, 00-716 Warsaw, Poland}
\author{Pierre Jean}
\affil{CESR, CNRS/Université Paul Sabatier Toulouse 3, BP 4346, 31028 Toulouse Cedex 4, France}
%\email{mariateresa.fiocchi@iasf-roma.inaf.it 
%}
%angela.bazzano@rm.iasf.cnr.it, pietro.ubertini@rm.iasf.cnr.it,
%aaz@camk.edu.pl
%% Notice that each of these authors has alternate affiliations, which
%% are identified by the \altaffilmark after each name.  Specify alternate
%% affiliation information with \altaffiltext, with one command per each
%% affiliation.
%% Mark off your abstract in the ``abstract'' environment. In the manuscript
%% style, abstract will output a Received/Accepted line after the
%% title and affiliation information. No date will appear since the author
%% does not have this information. The dates will be filled in by the
%% editorial office after submission.

\begin{abstract}
We report on the spectral analysis results of the neutron star, 
atoll type, low mass X-ray 
Binary \source observed
 by \integral and \sax satellites.
Spectral behavior in three different epochs 
corresponding to three different spectral states
has been deeply investigated.
Two data set spectra show a continuum well described by one or two soft 
blackbody plus a Comptonized
components with
changes in the Comptonizing electrons and
black body temperature and the accretion rates, 
which are typical of the spectral transitions
from high to low state.
In one occasion \integral spectrum shows, for first time in this source, 
a hard tail dominating the emission above 30 keV.
The total spectrum is fitted as the sum of 
a Comptonized component similar to soft state
and a power-law component ($\Gamma=2.76$), indicating the presence of a 
non thermal electron distribution of velocities.
In this case, a comparison with hard tails detected 
in soft states from neutron stars systems 
and some black hole 
binaries suggests that a similar mechanism could originate these components
in both cases.
\end{abstract}


\section{Introduction}
\source is a neutron star low mass X-ray binary (LMXB) 
classified as a atoll source (Hasinger \& van der Klis 1989), 
with an orbital period of 3.8 hr derived from the optical
variability of its companion V801 Arae (Pedersen, van Paradijs \& Lewin 1984)
and at distance of 3.7--6.5 kpc (Fujimoto et al. 1988, 
Smale \& Lochner 1992, Augusteijn et al. 1998).

While the X-ray burst properties and timing signatures have been analyzed extensively 
(see Jonker et al. 2005, Belloni et al. 2005 and references therein)
the spectral characteristics have been studied only at low energy with 
\einstein\/, \exosat\/, \temna\/ and \asca\/.
In general, the spectrum 
was acceptably fitted by a Comptonization model plus a black body component.
A coronal temperature of 2.3-­2.6 keV,
an optical depth of 13, and a soft black body temperature of 0.55 keV
were the best fit parameter for the Einstein data 
(Christian and Swank 1997).
\exosat\/ data reveal two different source intensity: 
%Vacca et al. (1987) chose two sections of their \exosat data as representative of
%periods of maximum and minimum source intensity and fitted them separately.
for the maximum intensity the coronal temperature was $\sim 3.5$ keV
and the optical depth $\sim 10.4$,
while for the minimum intensity the temperature was $\sim 6.1$ keV
and the optical depth $\sim 7.2$.
White et al. (1988) found the coronal 
temperature to be 1.8 keV and the optical
 depth to be 16.7. 
A model consisting of a black body,
a multi-color disk and a broad Fe line,
was used by Asai et al. (1998) to fit the \asca data, 
confirming previous \temna results (Waki et al. 1984).
We report here a broad band spectral analysis 
performed on data from \sax and \integral satellites, which allowed us to better 
constrain the spectral parameters and to detect the presence of a high energy tail
dominating the spectrum above $\sim 30$ keV.
A similar feature has been observed in other LMXBs namely GX~17+2 
(Di Salvo et al. 2000), 
GX~349+2 (Di Salvo et al. 2001), 
Sco~X-1 (D'Amico et al. 2001),
4U~1608-522 (Zhang et al. 1996), XB~1254-690 (Iaria et al. 2001), 
Cir~X-1 (Iaria et al. 2002) and 4U~0614+091 (Piraino et al. 1999).
\section{Observations and Data Analysis}
\label{observations}
Table \ref{jou}  summarizes the log of \integral and \sax observations of \source.

\sax observed the source on three occasions: February and March 1998 and
February 2000.
LECS, MECS and PDS event files and spectra,
available from the ASI Scientific Data Center,
were generated by means of the Supervised Standard Science Analysis
\cite{fio99}.
Both LECS and MECS spectra were accumulated in circular regions
of 8' radius. 
The PDS spectra were obtained
with the background rejection method based on fixed rise time thresholds.
Publicly available matrices were used for all the instruments.
The cross-calibration constant
values were taken in agreement with the indications given in Fiore et al. 1999.
Fits are performed in the following energy band: 0.5--3.5 keV for LECS, 1.5--10.0 
keV for MECS and 15--70 keV for PDS.\\
The analyzed \integral (Winkler et al. 2003) data set consists of all observations in which
4U~1636-53 was
within the high-energy detectors field of view.
Observations are organized into uninterrupted 2000 s long science pointing, windows 
(scw):
light curves and spectra are extracted for each individual scw.
Wideband spectra (from 5 to 150 keV) of the source are obtained using data from the
two high-energy
instruments JEM-X \citep{lun03} and IBIS \citep{ube03}.
Data were processed using the Off-line Scientific Analysis
(OSA version 5.1)
software released by the \integral Scientific Data Centre.
While IBIS  provide a very large FOV ($\sim 30\degr$), JEM-X has
a narrower FOV ($\sim 10\degr$), thus providing only a partial overlap with the high-energy
detectors.
Data from the Fully Coded field of view only for both instrument have been used.
The angular resolution of IBIS instrument is 12 arcmin, 
any source at a
distance larger than the instrument angular resolution do not contribute at all
to the observed source spectrum and flux (Ubertini et al. 2003)
due to the coded mask intrinsic characteristics.
\section{Spectral Analysis Results}
\label{analisis}
\begin{table}
\centering
\caption{BeppoSAX and INTEGRAL Observations}
\label{jou}
\scriptsize
\begin{tabular}{lccccccc}
\hline\hline
%& & & & & & &\\
\multicolumn{8}{c}{BeppoSAX Journal}\\
%& & & & & & &\\
& {\bf Start Date}&&{\bf Exposure time}&&&{\bf Count s$^{-1}$}&\\
& & & & & & &\\
&&LECS&MECS&PDS&LECS&MECS&PDS\\
&&ksec&ksec&ksec&[0.4-3 keV]&[1.5-10 keV]&[20-60 keV]\\
& & & & & & &\\
\emph{$1^{st}$ epoch (a)} &1998-02-24 &13 &39 &17 &$14.66\pm0.03$  &$39.37\pm0.02$$^{a}$ &$0.48
\pm0.04$\\
\emph{$1^{st}$ epoch (b)} &1998-03-01 &6  &14 &7  &$14.95\pm0.05$  &$25.42\pm0.04$$^{b}$ &$0.80
\pm0.06$\\
\emph{$1^{st}$ epoch (c)} &2000-02-15 &12 &37 &19 &$22.08\pm0.04$  &$29.09\pm0.03$$^{b}$ &$0.73
\pm0.03$\\
%& & & & & & &\\
\hline
%& & & & & & &\\
\multicolumn{8}{c}{INTEGRAL Journal}\\
& & & & & & &\\
&&{\bf Start Date}&\multicolumn{2}{r}{\bf Exposure time}&\multicolumn{2}{c}{\bf Count s$^{-1}$}
&\\
&&&JEM-X&IBIS&JEM-X&IBIS&\\
&&&ksec&ksec&[5-15keV]&[20-150 keV]&\\
& & & & & &&\\
%& & & & & &&\\
&\emph{$2^{nd}$ epoch} &2003-03-04  &36          &594         &$6.3\pm0.2$   &$10.74\pm0.08$&\\
&\emph{$3^{rd}$ epoch} &2003-03-04  &16          &117         &$3.70\pm0.06$ &$3.22\pm0.04$&\\
%&& & & & &&\\
\hline
\hline
\end{tabular}\\
{
$^{a}$ MECS count rates refer to MECS2 and MECS3 units.\\
$^{b}$ MECS count rates refer to only MECS2 unit.\\
}
\end{table}
The whole
data set was carefully fitted with several physical models,
while trying to keep the number of free parameters as low as possible. \\
Each time a new component was added to the model, a F-test was performed. We 
assumed that a F probability
larger than $95\%$ implies a significative improvement of the fit.
The uncertainties are at $90\%$ confidence level for one 
parameter of interest ($\Delta\chi^2=2.71$).\\
When spectra are from more than one detector, 
we allow the relative normalization to be free with respect to the 
MECS and IBIS data, for \sax\/ and \integral\/ respectively.
XSPEC v.\ 11.3.1. has been used.\\
Spectral behavior has been studied separately in three epochs consisting of the following data:

%\begin{itemize}
%\item
\emph{$1^{st}$ epoch}: all three \sax observations available from February 1998 
to February 2000. 
During these periods the source was always in a soft/high state.

%\item
\emph{$2^{nd}$ epoch}: JEM-X and IBIS data available from 52644 MJD to 
53644 MJD with count rate $<5~counts~s^{-1}$ in the 20--40 keV energy band. For the chosen period
the source was in a hard/low state.

%\item
\emph{$3^{rd}$ epoch}: JEM-X and IBIS data available from 52644 MJD to 53644 MJD 
with count rate $>5~counts~s^{-1}$ in the 20--40 keV energy band.
This epoch does not corrspond to either the soft or the hard state and  
and here we call it
\textit{peculiar} state as will be explained in detail later on.
%\end{itemize}

\begin{figure}[h]
\centering
\includegraphics[angle=-90,width=9.5cm]{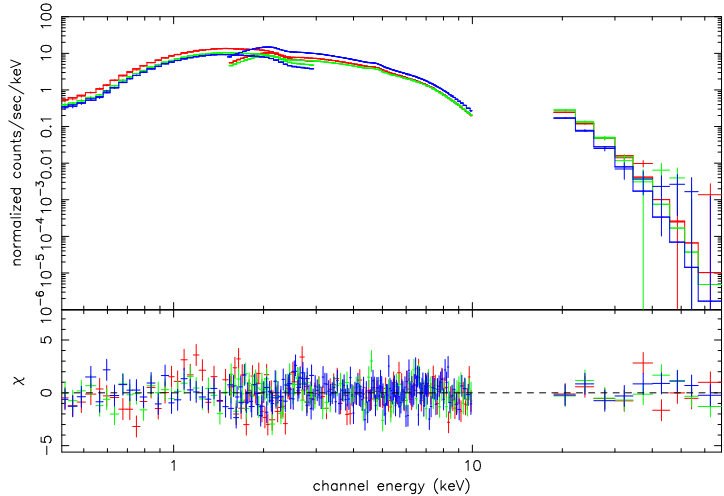}
\caption{
Three \sax spectra of \source and the
residuals with respect to the corresponding best fits, in the range 0.5-70 keV.
}
\label{saxs}
\end{figure}

The most simple model which provides a good fit to each \sax spectrum
in the energy band 0.5--70 keV
consist of a thermal Comptonized component modeled in XSPEC by
{\scriptsize{COMPTT}}
(a spherical geometry was assumed) plus
a soft component which we modeled with two temperature blackbody.
The simplest model consisting of multicolor  
{\scriptsize{DISKBB}} ~(Makishima et al. 1986) plus a {\scriptsize{COMPTT}} 
component do not give a good fit, with a $\chi^2_{rid}~\sim~2.5$ for each 
observations.
The black body and thermal Comptonized component parameters
were left free in order to determine the blackbody temperature $T_{bb}$,
the electron temperature $T_{e}$, the optical depth $\tau$ and seed photon
temperature $T_{o}$.
Results from these fits are reported in Table
\ref{fitsax}. Figure \ref{saxs} shows three \sax spectra and the 
residuals with respect to the corresponding best fits.
The column density {N$_{\rm H}$} towards the source
was left free and its value measured by the LECS and MECS instrument
is always close to the galactic column density 
($N_{\rm H}~galactic~=~3.58\times10^{21}~cm^{-2}$,
estimated from the 21 cm measurement of Dickey \& Lockman 1990).
The seed photon
temperature is always $T_{o}$=$1.3\pm0.2$ keV,
and all parameters did not significantly vary from one observation
to another.

\begin{figure}[h]
\centering
\includegraphics[angle=-90,width=9.5cm]{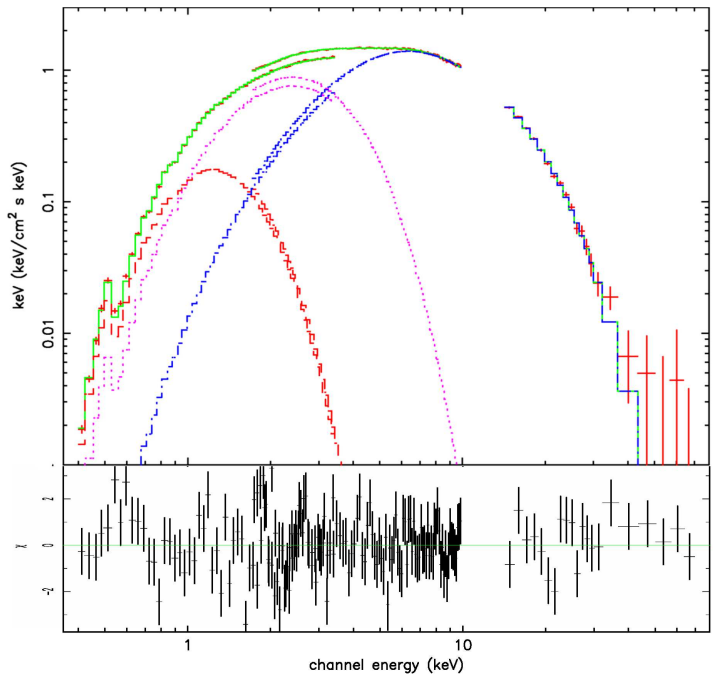}
\caption{The \sax average spectrum in the soft state (epochs 1), shown together with
the total model and its components: the total, two blackbody and the Comptonization
components are shown in green, red, magenta and blue, respectively.
Residuals with respect to the corresponding best fits are also showed.}
\label{sax1}
\end{figure}

In order to
achieve the highest signal to noise ratio
we build the \sax average spectrum arranging
all three observations.
This procedure can,
in principle, be risky since the source can change its
spectrum from one observation to another. However in our case,
the previous analysis showed no significant shape changes for the three observations.
We then can
take advantage of the high quality of the average spectrum up to about 70 keV.
We fit the average \sax spectrum with the model used for the single observations.
{N$_{\rm H}$} has been fixed to the value for the galactic column density.
%Being the column density {N$_{\rm H}$} towards the source 
%always close to the galactic column density for each \sax observation,
%we fix this parameter at this value.
Spectral fit results are given in Table \ref{fit}, the average spectrum is shown in 
Fig.\ \ref{sax1}.
\begin{table}
\scriptsize
\caption{Results of the fit of \source spectra in the energy band $0.5-70$ keV
for three \sax observations.
For each observations we report the best fit model.
The black body equivalent radius is computed using a distance of 5.9~kpc (Cornelisse et al. 2003).
Uncertainties are at the 90\% confidence level for a single parameter variation.}
\label{fitsax}
\centering
\begin{tabular}{ccccccccc}
\hline\hline
%&& & & & & & &\\
\multicolumn{9}{c}{\bf{BeppoSAX spectra}}\\
\multicolumn{9}{c}{$model=bbody+bbody+comptt$}\\
{N$_{\rm H}$} &{T$_{\rm BB1}$} & {T$_{\rm BB2}$}& {T$_{\rm e}$} &$\tau$ & $R_{\rm BB1}$& $R_{\rm BB2}$& $n_{\rm COMPTT}$ & $\chi^2$/d.o.f \\
$10^{21}~cm^{-2}$ &$keV$ &$keV$& $keV$ &&km &km& &\\
& & & & & & &\\
$3.1^{+0.5}_{-0.4}$&$0.28\pm0.03$&$0.61\pm0.03$         &$3.4^{+0.6}_{-0.5}$ &$3^{+2}_{-1}$ &$67^{+21}_{-5}$  &$24^{+2}_{-2}$ &$0.19^{+0.03}_{-0.02}$&190/124\\
$3.4^{+0.6}_{-0.3}$&$0.26\pm0.04$&$0.59^{+0.03}_{-0.04}$&$2.9^{+0.6}_{-0.3}$ &$5^{+2}_{-3}$ &$81^{+49}_{-12}$ &$23^{+3}_{-3}$ &$0.24^{+0.03}_{-0.05}$&132/122\\
$3.2^{+0.4}_{-0.2}$&$0.28\pm0.03$&$0.60\pm0.03$&$2.8^{+0.6}_{-0.4}$          &$5\pm2$       &$68^{+18}_{-13}$ &$21\pm2$        &$0.16^{+0.02}_{-0.04}$&133/122\\
%&& & & & & & &\\
\hline
\hline
\end{tabular}\\
\end{table}

The source was in the soft/high state
with an un-absorbed luminosity of $L_{0.1-200\,\rm keV} \simeq 2.0 \times 10^{37}$ erg s$^{-1}$, assuming a distance of 5.9 kpc (Cornelisse et al. 2003). 
As in the case of the \einstein observation (Christian and Swank 1997), 
thermal Comptonization of the
optically thick plasma corona with a quite low electron temperature
is dominating.\\
%The source energetics is dominated by the thermal Comptonization of the
%optically thick plasma corona with a quite low electron temperature,
%similar to state observed on \einstein data by Christian and Swank (1997),
%without any significant 
%hard X-ray emission.\\ 
The most simple model which provides a good fit to the \integral\/ hard state
consist of a thermal Comptonized component modeled in XSPEC by
{\scriptsize{COMPTT}} ~(Titarchuk 1994)
(a spherical geometry was assumed) plus
a soft component which we modeled by a single temperature blackbody.
Because of the very good low energy \sax coverage, we used 
the average value of input soft photon temperature and column density
($T_0=1.3 keV$ and $N_H=N_H~Galactic $) measured by the
BeppoSAX observations.
During this period, the source has a 
luminosity of $L_{0.1-200\,\rm keV} \simeq 1.4 \times 10^{37}$ erg s$^{-1}$, 
lower than the one in the soft state, 
in agreement with the usual ranking of the luminosity 
in atoll sources 
(e.g., Hasinger \& van der Klis 1989; van der Klis 2000; Gierli\'nski \& Done 2002).
 The electron temperature is now substantially higher than in the soft state, 
$T_{\rm e}\sim 23$ keV, and the Comptonization component extends well above $\sim 100$ keV.
\begin{table}
\scriptsize
\caption{Results of the fit of \source spectra in the energy band $0.5-70$ keV and $5-150$ keV
for \sax and \integral observations respectively, for three different spectral states.
For each state we report the best fit model.
The black body equivalent radius is computed using a distance of 5.9~kpc (Cornelisse et al. 2003).
Uncertainties are at the 90\% confidence level for a single parameter variation.}
\label{fit}
\centering
\begin{tabular}{ccccccccc}
\hline\hline
%&& & & & & & &\\
\multicolumn{9}{c}{\bf{BeppoSAX average high/soft state spectrum}}\\
\multicolumn{9}{c}{$model=bbody+bbody+comptt$}\\
{T$_{\rm BB1}$} & {T$_{\rm BB2}$}& $T_{0}$  &{T$_{\rm e}$} &$\tau$ & $R_{\rm BB1}$& $R_{\rm BB2}$& $n_{\rm COMPTT}$ & $\chi^2$/d.o.f \\
$keV$ &$keV$ &$keV$& $keV$ &&km &km& &\\
& & & & & & &\\
$0.24\pm0.01$&$0.58\pm0.02$&$1.3\pm0.1$ &$3.4\pm0.3$ &$3.8\pm0.4$ &$79\pm6$ &$25\pm1$ &$0.19\pm0.02$&235/190\\
%&& & & & & & &\\
\hline
%& & & & & & && \\
\multicolumn{9}{c}{\bf{INTEGRAL low/hard state spectrum}}\\
\multicolumn{9}{c}{$model=bbody+comptt$}\\
&{T$_{\rm BB2}$}& $T_{0}$  &{T$_{\rm e}$} &$\tau$ & $R_{\rm BB2}$ & $n_{\rm COMPTT}$ & $\chi^2$/d.o.f& \\
&$keV$ &$keV$& $keV$ &&km &$10^{-2}$ &&\\
& & & & & & &\\
&$1.2_{-0.3}^{+0.2}$&1.3~fixed &$23_{-2}^{+7}$&$1.1_{-0.3}^{+0.5}$&$5^{+3}_{-2}$ &$1.0^{+0.3}_{-0.1}$ &57/59&\\
%& & & & & & &&\\
\hline
%& & & & & & &&\\
\multicolumn{9}{c}{\bf{INTEGRAL peculiar state spectrum}}\\
\multicolumn{9}{c}{$model=comptt+powerlaw$}\\
&$T_{0}$  &{T$_{\rm e}$} &$\tau$ &$\Gamma$ & $n_{\rm COMPTT}$ & $n_{\rm powerlaw}$&$\chi^2$/d.o.f& \\
&$keV$&$keV$& &&$10^{-2}$ &$ph~keV^{-1}~cm^{-2}$&&\\
& & & & & & & &\\
&1.3~fixed&$5^{+4}_{-3}$ &$3\pm2$&$2.6\pm0.1$&$2.8_{-0.6}^{+0.8}$ &$1.0^{+1.4}_{-0.5}$ &72/57&\\
%& & & & & & & &\\
\hline
\hline
\end{tabular}\\
%$^a$ Parameters are in the range of BeppoSAX
\end{table}

\begin{figure}
\centering
\includegraphics[angle=-90,width=7.6cm]{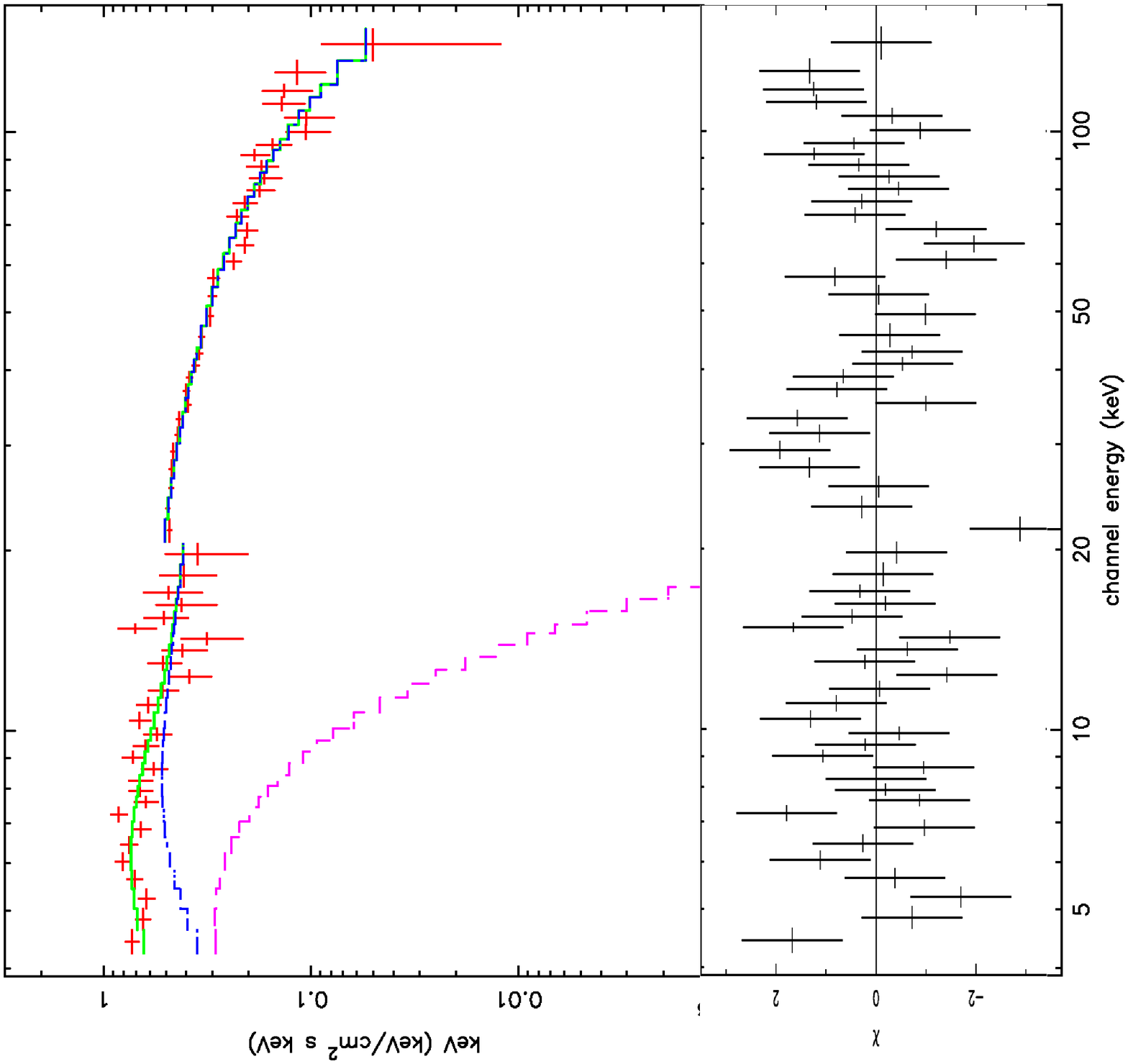}
\includegraphics[angle=-90,width=7.6cm]{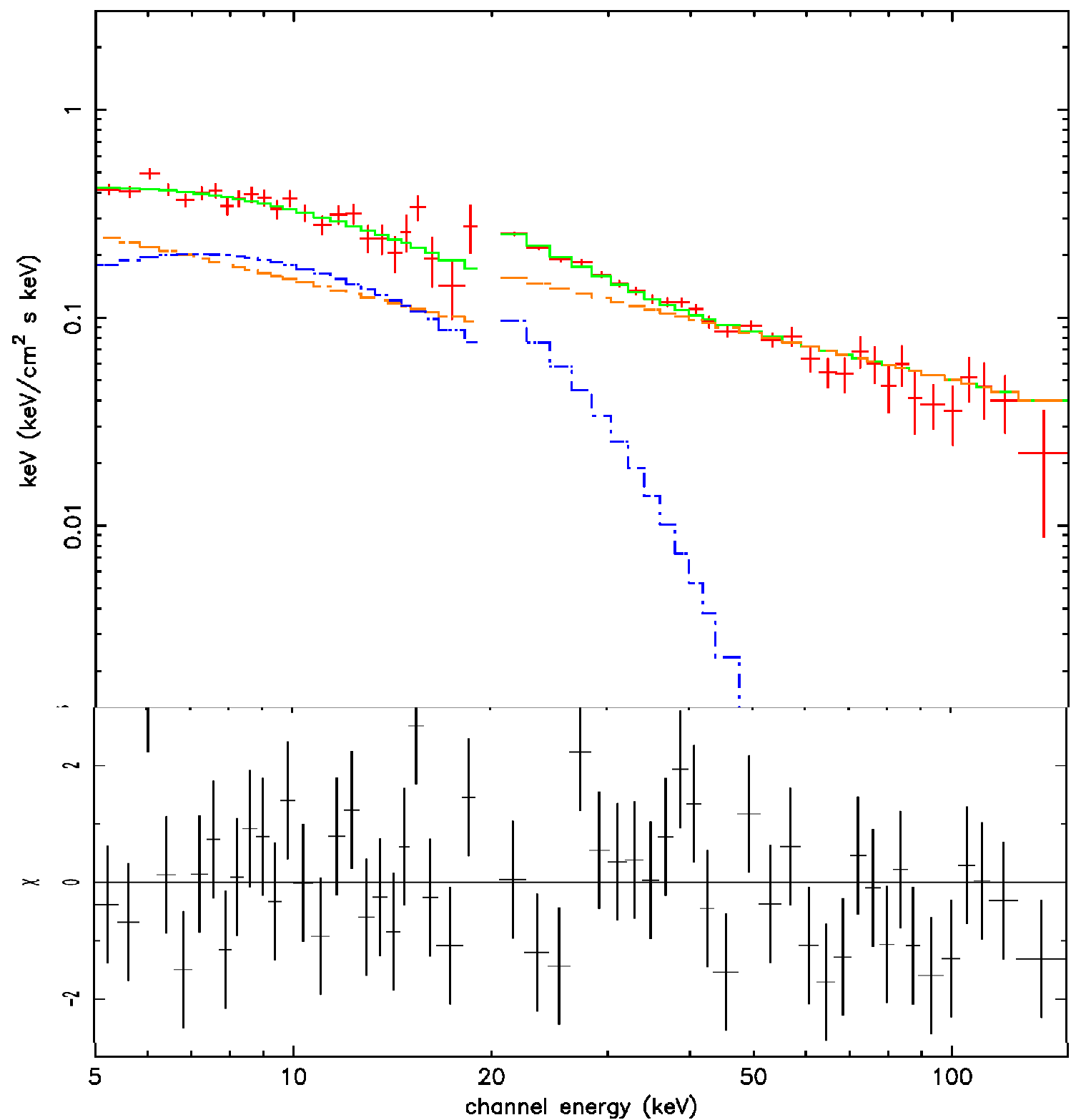}
\caption{The spectra of epochs 2 (soft state, left) and 3 (\textit{peculiar} state, right)
observed by \integral, shown together with the total model and its components. Left: the total,
the blackbody and the Comptonization components are shown in green, magenta and blue, respectively.
Right: the total, the Comptonization and the power law component are shown in green,
blue and orange respectively. Residuals with respect to the corresponding best fits are also showed.}
\label{int1}
\end{figure}

\begin{figure}
\centering
\includegraphics[angle=-90,width=9.5cm]{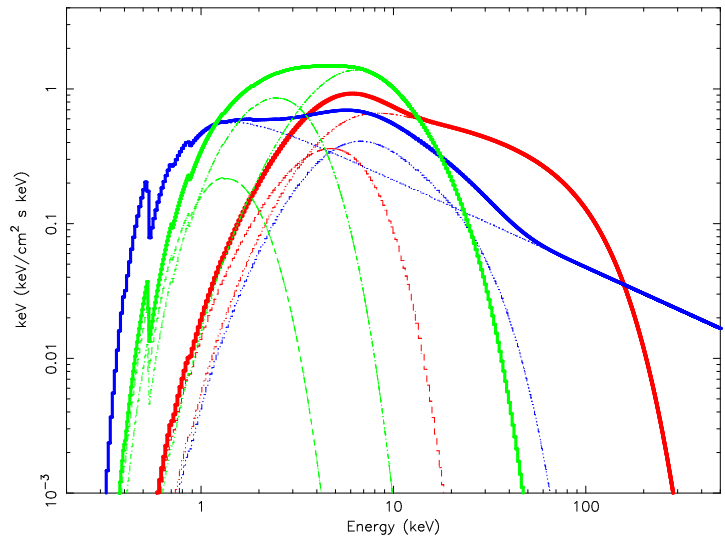}
\caption{
Comparison of the models for the three observations. Epoch 1, \sax soft state: green; epoch 2, \integral hard state: red; epoch 3, \textit{peculiar} \integral state: blue.
}
\label{models}
\end{figure}

The same model was been applied to the 
$2^{nd}$ \integral\/ data set but resulted in a poor fit with 
a $\chi^2/d.o.f=97/57$ and clear residuals above 60 keV.
Adding a power law improves the fit significantly 
($\chi^2/d.o.f$ 
becomes $64/55$).
A simple model with the single temperature blackbody and power law
do not fit our data ($\chi^2_{rid}~\sim~5$).
The disk component becomes negligible and it is not necessary to best fit the data.
Spectral fit
 results are given in Table \ref{fit}, and spectra are shown in Fig.\ \ref{int1}
.
In this \textit{peculiar} state,
the spectrum is also compatible with a power-law and multiple blackbody (Makishima et al. 1986)
instead of Comptonization halo;
nevertheless the temperature is extremely high and perhaps unphysical.
Since the spectral parameters in the \textit{peculiar} state are similar 
to those in the soft/high \sax state, a possible interpretation of this state is
that \source was in a similar soft/high state 
as during the \sax observations, 
but with a new overlapped component at high energies simultaneously present.
The un-absorbed luminosity is $L_{0.1-200\,\rm keV} \simeq 7.4 \times 10^{37}$ erg s$^{-1}$.
Comparison between the models in the three different 
spectral states are shown in Fig. \ref{models}.
Spectral state transitions are evident, with the hard INTEGRAL state 
extended up well above 100 keV and, 
in the \textit{peculiar} state, a steep power-law 
component detected with high statistical significance up to 100 keV.\\ 

\section{Discussion and Conclusions}
\label{discussion}
According to our present understanding,
the black-body component in the soft states could originate at both the 
neutron star surface and the surface of an optically-thick accretion disk. 
In our observations the two black body components seems to originate          
from two different parts of the disk, 
corrisponding to two different temperature.
The high optical thickness of the plasma
rules out the black body component as originates from the neutron star cap regions.
Anyway, independently from the optical thickness of the plasma,
it is very
improbable to have emission from the polar caps of the neutron
star for this source. Infact the magnetic field of the LMXBs is $\sim$
$10^8$ Gauss implying that the accretion onto the neutron  star is
not magnetically dominated and the matter accretes onto the whole
NS surface.
The Comptonization component may arise from a corona above the disk and/or 
between the disk and the neutron star surface. 
In the hard states, accretion probably assumes the form of a truncated 
outer accretion disk and a hot inner flow, 
joining the disk and the stellar surface as previously reported
by Barret \& Olive (2002) for LMXB 4U~1705-44.
The spectral transitions are generally, but not necessarily, 
coupled with changes in luminosity, 
indicating they are driven by variability of the accretion rate 
or change of the geometry of the system as for Black Hole hard/soft transition 
at constant luminosity (Belloni et al. 2005).
For \source the accretion rate is lower in the hard state than in the soft state:
for an accretion efficiency of $\eta = 0.2$ 
(corresponding, e.g., to $M_{\rm NS}=1.4 {\rm M}_{\sun}$ and $R_{\rm NS}=10$ km ) 
and using our model luminosities, $L_{0.1-200\,\rm keV}$,
we get 
$\dot{M}_{soft}\simeq 7.3 \times 10^{-9} {\rm M}_{\sun}$ yr$^{-1}$ and  
$\dot{M}_{hard}\simeq 4.0 \times 10^{-9} {\rm M}_{\sun}$ yr$^{-1}$.    
Transitions are apparently accompanied by changing in 
the geometry of the flow, 
and in the relative contribution of the blackbody-like and Comptonization components. 
In the peculiar state the accretion rate becames very high 
($\dot{M}_{peculiar}\simeq 2.1 \times 10^{-8} {\rm M}_{\sun}$ yr$^{-1}$),
but this value can be influenced by the steep power law
that dominates the
energy spectrum at low energies.
The emission is best described 
by Comptonization from a complex
electron distribution due to a low temperature
($\sim 3-4$keV) thermal electron distribution 
%({\.Z}ycki, Done
%\& Smith 2001; Kubota, Makishima \& Ebisawa 2001) 
together with
non-thermal power-law electrons. 
%(Gierli{\'n}ski et al.~1999;
%Zdziarski et al.~2002). 
This two-component electron distribution
could arise from non-thermal electron
acceleration regions powered by magnetic reconnections above a
disc. Low-energy electrons cool preferentially by Coulomb
collisions leading to a thermal distribution while the
high-energy electrons cool by Compton scattering, preserving a
non-thermal distribution (Coppi 1999). Alternatively, the
thermal and non-thermal electrons could be spatially distinct,
e.g.\ magnetic reconnection above the disc can produce a
non-thermal electron distribution, while overheating of the
inner disc produces the thermal Comptonization (Kubota et
al.~2001).

LMXBs exhibit hard X-ray states which are very 
similar to the Black Hole Candidates ones and 
investigating the possible disc-jet coupling is very interesting
and still matter of discussion.
The power-law component could be produced by Componization by
synchrotron emission in the jet (Bosch-Ramon et al. 2005, Fender 2004) 
as proposed for others LMXBs: 
the power law component in Cir~X-1 is explained
by Iaria et al. (2002) as jet emission that has been resolved using 
radio interferometry by Fender \& Kuulkers (2001);
a significant correlation between radio and X-ray flux
has been detected for 4U~1728-34 (Migliari et al. 2003), indicating
a clear
signature of disc-jet coupling. 
In our source, this hypothesis is strengthened by radio detection:
Sydney University Molonglo Sky Survey catalog gives a flux of 7.5 mJy at 843 GHz
(Mauch et al. 2003).
Assuming a radio spectral index of -0.5 to estimate the flux density at 5 GHz 
based on other observations, we found a radio loudness $P_R/P_X$ (as defined in Fender \& Kuulkers 2001)
in the range 1.1-2.2 Jy/Crab, where $P_R$ is the radio flux density at 5 GHz 
and $P_X$ 
is the peak X-ray flux measured in the soft band ($<12~keV$), even though the
two data set are not simultaneous.
The radio loudness value is consistent with the one for Cir X-1 (Fender \& Kuulkers 2001) and
it is very high respect to other LMXBs and 
quite similar to that of Back Hole Candidates.
These founding are in favour of the jet hypothesis as origin of the power-law
observed for \source similarly to 
GX~5-1, Cir~X-1 and GX~17+2 being all of them associated with variable radio sources 
(Fender \& Hendry 2000, and reference therein).

Simultaneous high energy and radio observations during spectral
transition are crucial
to disentangle the power law emission in the LMXBs containing a neutron star.

Finally, Laurent \& Titarchuk (1999) suggest the detection of a power-law 
component at high energy 
to be a signature of presence black hole in an X-ray binary system. Our data
are supporting result on GX~17+2 by Di Salvo et al. (2000),
clearly showing this criterion proposed to distinguish black hole versus 
neutron star binaries is inadequate.
%\vspace{-0.7cm}
\begin{acknowledgments}
We acknowledge the ASI financial/programmatic support via contracts ASI-IR 
046/04.
A special thank to M. Federici and G. De Cesare for supervising 
the \integral data archive and software respectively.
A very particular thank to K. Kretschmer for making data 
available before becoming public.
\end{acknowledgments}

\end{document}